\documentclass{PoS}
\usepackage{amsmath}
\usepackage{multirow}

\title{Color-electric correlation functions under gradient flow}

\ShortTitle{Color-electric correlation functions under gradient flow}

\author{\speaker{Hai-Tao Shu}\\
        Fakult\"at f\"ur Physik, Universit\"at Bielefeld, 33615 Bielefeld, Germany\\
        E-mail: \email{htshu@physik.uni-bielefeld.de}}

\author{Luis Altenkort \\
        Fakult\"at f\"ur Physik, Universit\"at Bielefeld, 33615 Bielefeld, Germany\\
        E-mail: \email{
altenkort@physik.uni-bielefeld.de}}

\author{Olaf Kaczmarek \\
        Fakult\"at f\"ur Physik, Universit\"at Bielefeld, 33615 Bielefeld, Germany\\
Key Laboratory of Quark \& Lepton Physics (MOE) and Institute of Particle Physics,\\
        Central China Normal University, Wuhan 430079, China\\
        E-mail: \email{okacz@physik.uni-bielefeld.de}}

\author{Lukas Mazur \\
        Fakult\"at f\"ur Physik, Universit\"at Bielefeld, 33615 Bielefeld, Germany\\
        E-mail: \email{lmazur@physik.uni-bielefeld.de}}

\abstract{

We report on the progress of our study on the color-electric correlation functions under gradient flow on the lattice. This calculation is the first step of our long-term project to estimate a series of important transport coefficients, of which the heavy quark momentum diffusion coefficient is our first attempt, as it can be extracted from a color-electric correlation function that has been calculated non-perturbatively using noise reduction technique in the quenched approximation~\cite{Francis:2015daa}. By comparing the flowed correlation function with those obtained by other signal-improving techniques, for instance the multi-level algorithm used in~\cite{Francis:2015daa}, one can gain insight into the applicability of the gradient flow and the renormalization of the correlation functions.

We start with quenched, isotropic lattices. Currently we have finished measuring the color-electric correlation functions on 4 different lattices with $\beta$-values corresponding to a temperature of $T\approx 1.5T_c$. We perform a continuum extrapolation on the flowed correlators at fixed physical flow times followed by an extrapolation of the continuum estimate back to zero flow time. The next step is to extend the study to different temperatures and to extend the study to dynamical QCD.

}

\FullConference{37th International Symposium on Lattice Field Theory - Lattice2019\\
		16-22 June 2019\\
		Wuhan, China}

\begin{document}

\section{Introduction}

Gradient flow is a powerful tool to reduce noise when measuring quantities constructed from gauge fields. It can be proven that at positive flow time, the flowed gauge fields are smooth and operators are automatically renormalized if the continuum limit is taken~\cite{Luscher:2010iy}. This technique is particularly useful when dealing with high dimensional quantities, for instance the energy-momentum tensor or topological charge, as they are prone to noise caused by high frequency fluctuations of the gauge fields on the lattice. Methods like the multi-level algorithm~\cite{Luscher:2001up} and link-integration~\cite{DeForcrand:1985dr} can be used to improve the signal. However, they are only applicable in pure gauge theories. The approach of the gradient flow is still applicable even when dynamical quarks are introduced, which makes it the tool of choice for this work.

Some previous works that successfully utilize the gradient flow have already been published. For instance, by measuring the energy-momentum tensor and its correlation functions under flow, the equation of state, entropy density and specific heat can be obtained~\cite{Kitazawa:2016dsl,Kitazawa:2017qab}. Studies of correlators of topological charge density related to the sphaleron rate can be found in~\cite{Kotov:2018vyl,lukas}. In this work we measure the color-electric correlation function~\cite{CaronHuot:2009uh},
\begin{equation}
\label{gee}
G_{EE}(\tau)=-\frac{1}{3}\sum_{i=1}^3\frac{\langle {\rm{Re}\ \rm{Tr}}[U(\beta,\tau)gE_i(\tau,\vec{0})U(\tau,0)gE_i(0,\vec{0})]\rangle}{\langle {\rm{Re}\ \rm{Tr}}[U(\beta,0)]\rangle}
\end{equation}
under gradient flow, where $\beta$ is the temporal extent of the lattice, $U$ is a Wilson line and $gE_i$ is the color-electric field. We discretize Eq.\ref{gee} in the way proposed in \cite{CaronHuot:2009uh}. For the flow we use a Symanzik improved version~\cite{Ramos:2015baa}.

The gradient flow arises through the flowed SU(3) gauge field $B_{\mu}(\tau_F,x)$, which is defined by the flow equations
\begin{equation}
\label{flow_equation}
\dot{B}=D_{\nu}G_{\nu\mu},\ \ G_{\mu\nu}=\partial_{\mu}B_{\nu}-\partial_{\nu}B_{\mu}+[B_{\mu},B_{\nu}],\ \ D_{\mu}=\partial_{\mu}+[B_{\mu},\cdot],
\end{equation}
with the initial condition $B_{\nu}|_{\tau_F=0}=A_{\nu}$. Here $\dot{B}_\mu$ denotes the derivative of the flowed gauge field with respect to the flow time $\tau_F$. In $D$ dimensions the leading-order solution reads~\cite{Luscher:2010iy}    
\begin{equation}
\label{flow_solution}
B_{\mu}(\tau_F,x)=\int d^Dy\ K_{\tau_F}(x-y)A_{\mu}(y),\ \ K_{\tau_F}(z)=\int \frac{d^Dp}{(2\pi)^D} e^{ipz}e^{-\tau_F p^2}=\frac{e^{-z^2/4\tau_F}}{(4\pi \tau_F)^{D/2}},
\end{equation}
and reveals that the flow averages gauge fields around a gaussian envelope with a ``flow radius'' of $r_F\equiv \sqrt{8\tau_F}$. In order to ensure that lattice discretization effects are sufficiently suppressed, one requires $r_F \gtrsim a$. On the other hand the flow should not touch the lattice boundary, i.e. the flow radius should be smaller than half the temporal extent of the lattice $1/(2T)$. Another condition emerges from the fact that the correlation functions of two operators sitting at a distance of $\tau$ should be well separated in order to avoid overlapping. We therefore have a general condition for the flow time~\cite{Kitazawa:2017qab}
\begin{equation}
\label{flow_time_requirement}
a\lesssim \sqrt{8\tau_F} \lesssim \tau \leq \frac{1}{2T}.
\end{equation}
Specific to the the color-electric correlation function that we are considering in this work a leading order perturbative calculation suggests an upper limit~\cite{Eller:2018yje}
\begin{equation}
\label{flow_time_more_requirement}
\tau_F < 0.014 \tau^2 ,
\end{equation}
before the correlation function is strongly affected by the flow. 
In the following discussions we will use the definition of the rescaled dimensionless flow radius $r_F T= \sqrt{8\tau_F} T$ and flow time $\tau_F T^2$. Putting the above conditions together we obtain
\begin{equation}
\label{flow_time_full_limit}
\frac{1}{N_{\tau}} \lesssim r_FT \lesssim \frac{\tau T}{3}.
\end{equation}

The main idea of the flow technique is to smear the fields along an additional dimension $\tau_F$. The high-precision data for the correlator that is obtained at positive flow time has to be extrapolated to the continuum limit first and has to be extrapolated to zero flow time afterwards in order to go back to the original $4D$ spacetime. The flow time condition from above has to be respected for the $\tau_F\rightarrow 0$ extrapolation. In the next section we present the details of our calculations on the lattice.

\section{Lattice Setup}

\begin{table}[htb]
\centering
\begin{tabular}{ccccccc}                                                                                                                                      
\hline
$\beta$ & $a${[}fm{]}($a^{-1}${[}GeV{]}) & $N_{\sigma}$ & $N_{\tau}$ & $T/T_{c}$ & \#confs.\tabularnewline
\hline
\hline
\multirow{1}{*}{6.8736} & \multirow{1}{*}{0.026 (7.496)} & \multirow{1}{*}{64} & 16  & 1.5  & 10000 \tabularnewline
\hline
\multirow{1}{*}{7.0350} & \multirow{1}{*}{0.022 (9.119)} & \multirow{1}{*}{80} & 20  & 1.5  & 10000 \tabularnewline
\hline
\multirow{1}{*}{7.1920} & \multirow{1}{*}{0.018 (11.19)} & \multirow{1}{*}{96} & 24  & 1.5  & 10000 \tabularnewline
\hline
\multirow{1}{*}{7.3940} & \multirow{1}{*}{0.014 (14.21)} & \multirow{1}{*}{120} & 30  & 1.5  & 10000 \tabularnewline
\hline
\end{tabular}
\caption{ $\beta$ values, lattice spacings, lattice sizes and number of configurations
measured in this work.}
\label{lattice_setup}
\end{table}

We summarize the lattice setup of this work in table \ref{lattice_setup}. The lattice spacing $a$ is determined by Sommer parameter $r_0$ \cite{Sommer:1993ce}, where we use a parameterization from \cite{Sommer:1993ce} with updated coefficients from \cite{Burnier:2017bod}. For each lattice we flow to 10 discrete flow times in the range of $r_FT\in [0,0.1]$ using an adaptive step-size Runge-Kutta algorithm. We then measure the correlator on 10,000 configurations for each lattice. With such large statistics we found that on the finest lattice, the relative statistical error of the correlation function at the middle point at flow radius $r_FT=0.05$ reaches $\sim$1.5\%. The statistical errors are even smaller for smaller $\tau T$ or larger flow time. This allows us to perform reasonably accurate extrapolations.

\section{Results}

We normalize the measurements of the lattice correlator with
~\cite{CaronHuot:2009uh}
\begin{equation}
\label{gcont}
 G_\mathrm{norm}(\tau T)=\pi^2 T^4 \left[
 \frac{\cos^2(\pi \tau T)}{\sin^4(\pi \tau T)}
 +\frac{1}{3\sin^2(\pi \tau T)} \right],
\end{equation}
which helps to visualize the details of the correlator. Additionally, we correct the correlation distances $\tau T$ by employing tree-level improvement according to~\cite{Sommer:1993ce,Meyer:2009vj}.

\begin{figure}[h]
\centerline{
\includegraphics[width=0.5\textwidth]{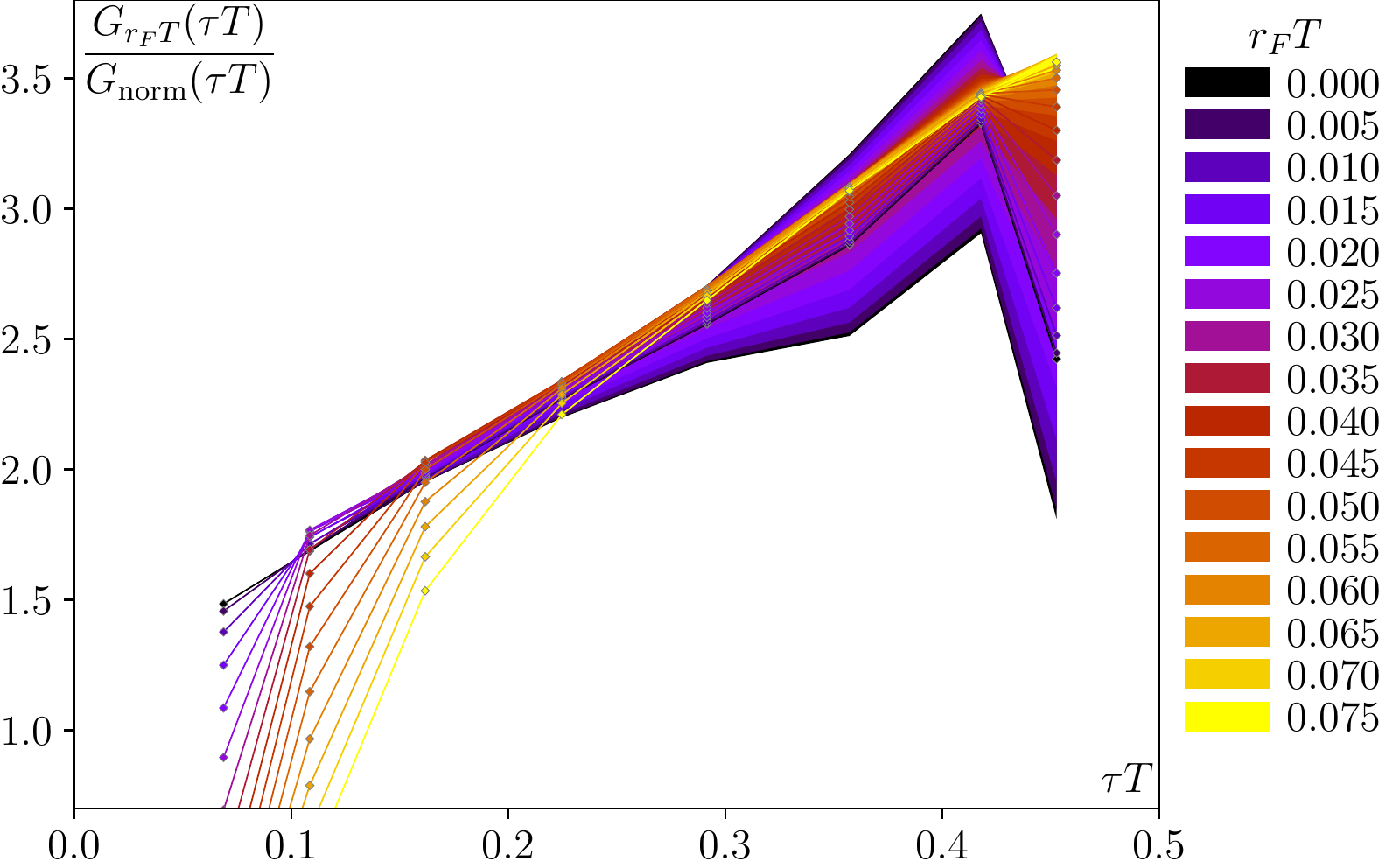}
\includegraphics[width=0.5\textwidth]{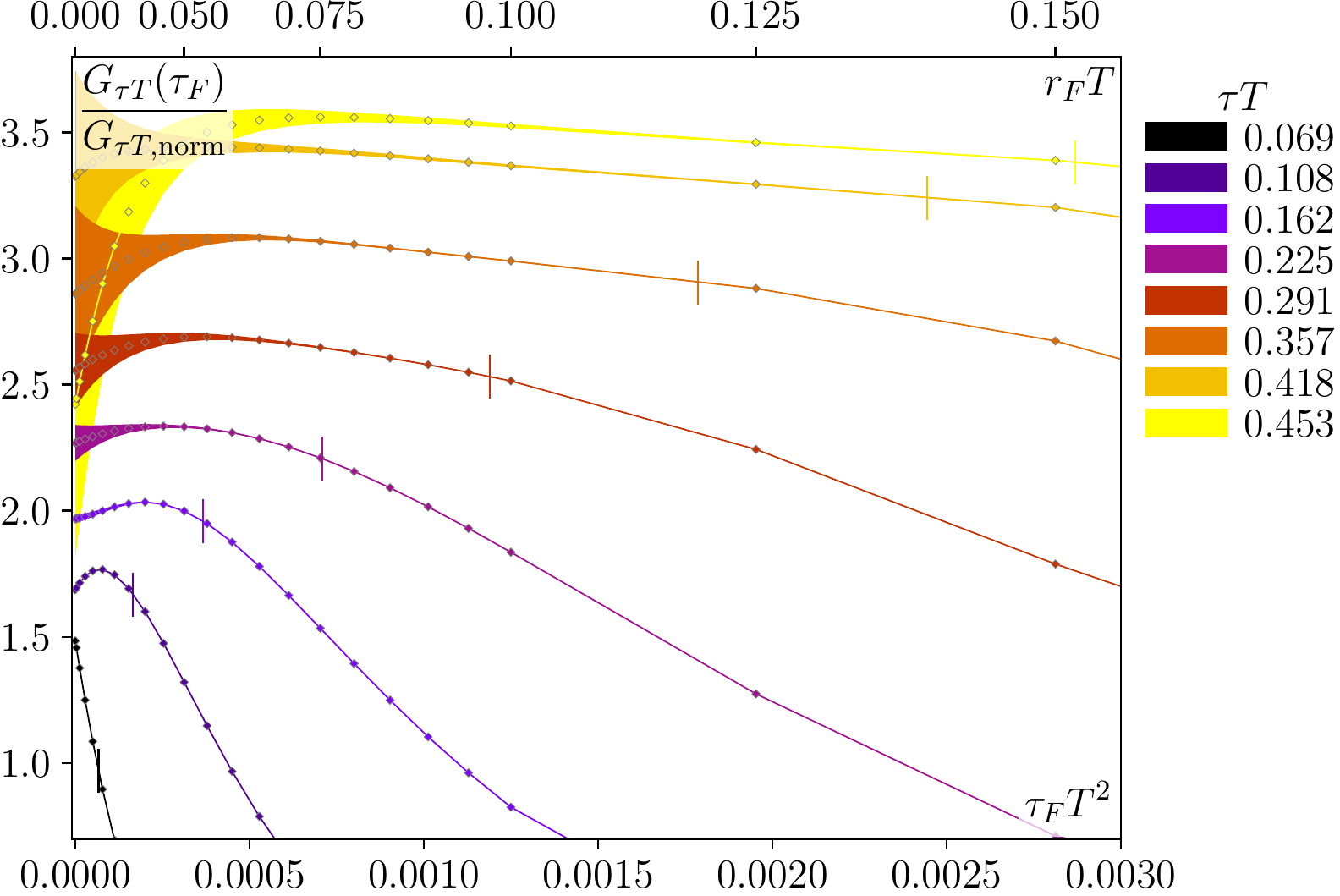}
}
\caption{\textit{Left}: Flow effects on the correlators observed on $64^3\times 16$ lattice. \textit{Right}: Upper limits of flow time for different $\tau T$ taken from~\cite{Eller:2018yje} on $64^3\times 16$ lattice, represented as vertical lines.}
\label{flow_effect2}
\end{figure}

Figure~\ref{flow_effect2} (left) shows the correlator at various fixed flow radii $r_FT$ as a function of distance $\tau T$. Here one can see that the larger the distance, the longer it takes for the flow to produce a reasonable signal. However, for the smallest distances the correlator quickly drifts towards zero, which suggests that the relevant fluctuations in which the actual physics lie in are destroyed immediately by the flow. 
For all $\tau T$ the statistical errors are improved a lot, for instance from $r_FT=0$ to $r_FT=0.075$, the relative error of the correlator at $\tau T=0.225$ is reduced from 3.15\% to 0.04\%, by a factor of $\approx 73$.
On the right panel of figure~\ref{flow_effect2} we show the correlator at fixed $\tau T$ as a function of flow radius. We also include the leading order perturbatively calculated flow time limits\cite{Eller:2018yje} as vertical lines. 
After a certain amount of flow, the operators do not ``feel'' the underlying lattice anymore and obtain their renormalized continuum values. After this effect we observe that for most $\tau T$ there is a range of flow times that respect the flow limit and are also large enough to produce high-precision signals. Within this range the flow has only a minor effect on the correlator, which we exploit to extrapolate it to zero flow time with a linear Ansatz.

\subsection{Continuum extrapolation}

In this section we show how we carry out the continuum extrapolation using a method that was developed in \cite{hauke_thesis}. As an example we consider the extrapolation for a single flow radius $r_FT$. For each lattice we select the correlator at this $r_FT$ and perform a combined spline fit, where the coeffcients of the spline depend on $N_\tau$. From that we simultaneously obtain the interpolation of the correlators and the continuum estimate ($a\rightarrow 0$ or $N_\tau \rightarrow \infty$). The splines are piece-wise defined polynomials of the general form
\begin{align}
\label{combine_spline}
G(\tau T)=\sum_{i=0}^{d} a_i\big{(}\tau T-s\big{)}^i+\sum_{j=0}^{n}c_j(\tau T-t_j)_{+}^d,
\end{align}
where $(x)_{+}$ is a step function that equals $x$ for positive $x$ and zero otherwise. $d$ is the degree of the polynomials and $n$ denotes the number of knots whose positions are $t_j$. $a_i$ and $c_j$ are coefficients to fit. 

\begin{figure}[h]
\centerline{
\includegraphics[width=0.52\textwidth]{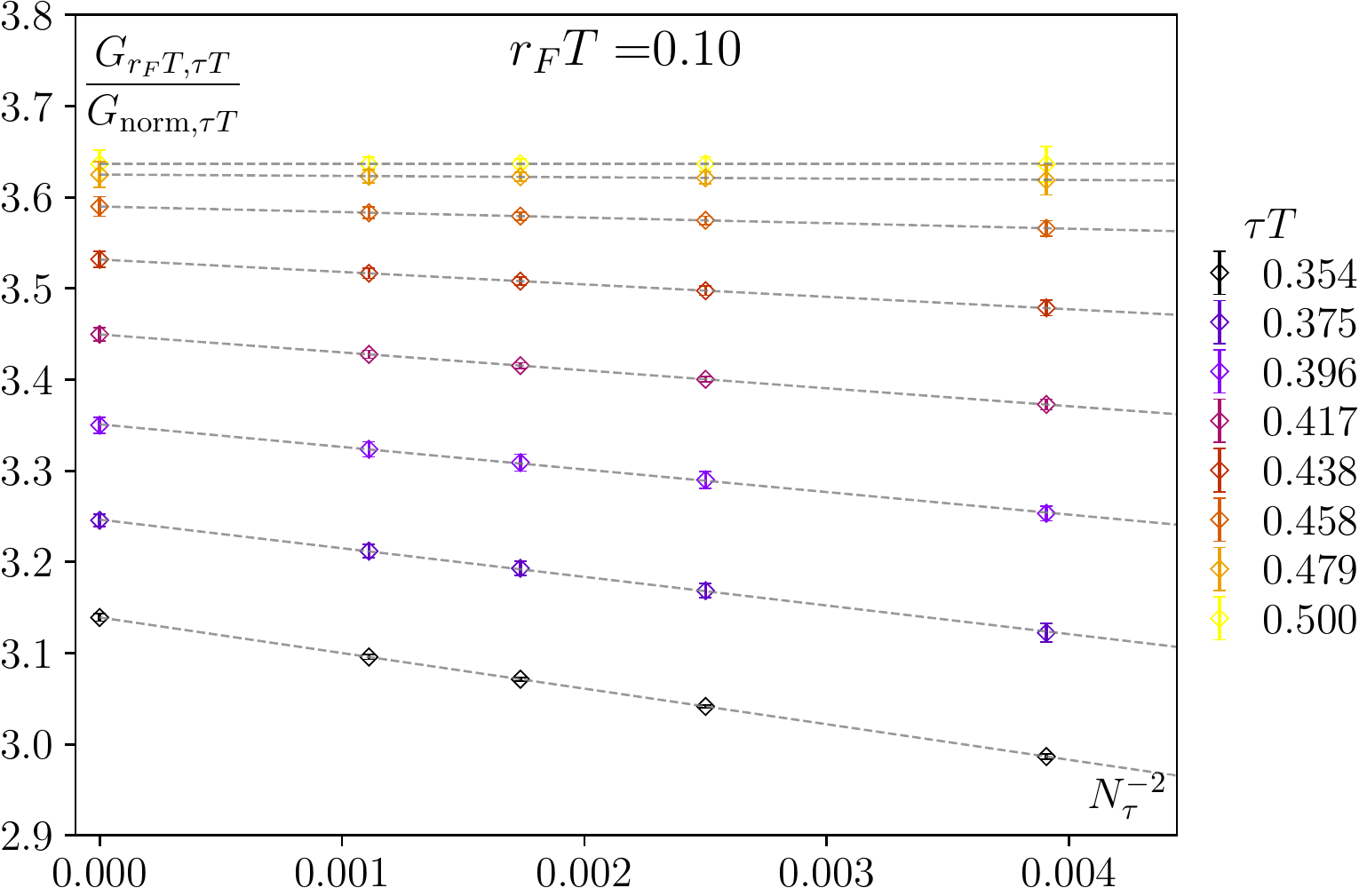}
\includegraphics[width=0.48\textwidth]{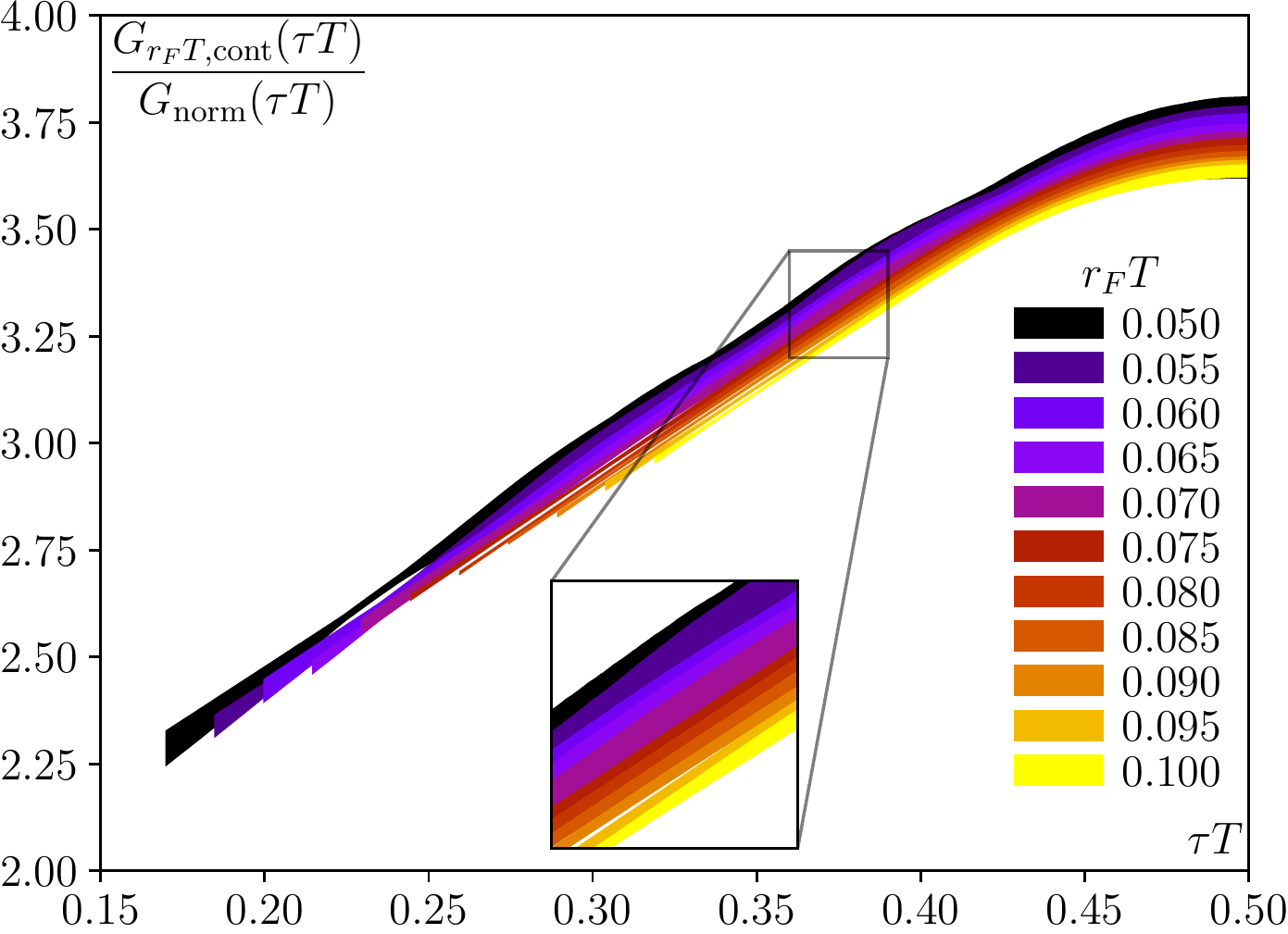}
}
\caption{\textit{Left}: Continuum extrapolation of the correlators using $1/N_{\tau}^2+b$ ansatz at $r_FT=0.10$. \textit{Right}: Continuum-extrapolated correlators for different flow times.}
\label{cont_extr}
\end{figure}

The two terms on the right hand side of equation~\ref{combine_spline} play different roles. The first term takes care of the continuum extrapolation and the second term is responsible for the interpolation to the same $\tau T$. Here we point out that the coefficients $a_i$ are not just numbers but functions, whose form is determined by the operator considered. In this work we choose a customary Ansatz for actions with $\mathcal{O}(a^2)$ cutoff-effects, namely
\begin{align}
a_i=\frac{m}{N_\tau^2}+b,
\end{align}
where $m$ and $b$ are fit parameters. For $1/N^2_{\tau}\rightarrow 0$ one then obtains the corresponding coefficients for the continuum-extrapolated correlator. In the left panel of figure~\ref{cont_extr} we illustrate this procedure for some selected $\tau T$ at $r_FT=0.10$. We then repeat it for all the flow times and summarize all extrapolation in the right panel of figure~\ref{cont_extr}. 

Note that the parameters of the spline have to be tuned by hand and are chosen such that the average $\chi^2/\mathrm{d.o.f.}$ is as close to $1$ as possible. We find that $d=2$ and averaging over $n\in 1,2,3$, where we distribute the knots evenly, produces the best results. We also constrain the first derivative of the splines to be zero at $\tau T=0.5$, as the correlator is symmetric because of the periodic boundary conditions. Additionally, we decide to constrain the second derivative of the spline to be zero at $\tau T =0$. This constraint far outside the fitting interval greatly stabilizes the extrapolations across different flow times for the intermediate $\tau T$. With our current lattice setup we can not explore the lowest $\tau T$, where one should then not use this constraint anymore. Only data points that respect equation \ref{flow_time_full_limit} are taken into account for the fit. Besides we set the minimal flow radius to $r_FT=0.05$ instead of taking the exact lower limit of equation \ref{flow_time_full_limit} as the latter would remove many data points for the lower $\tau T$. This ensures that we have a usable signal everywhere. 
One extrapolation is visualized in the left panel of figure ~\ref{cont_extr}.

\subsection{Zero flow time extrapolation}

\begin{figure}[h]
\centerline{
\includegraphics[width=0.5\textwidth]{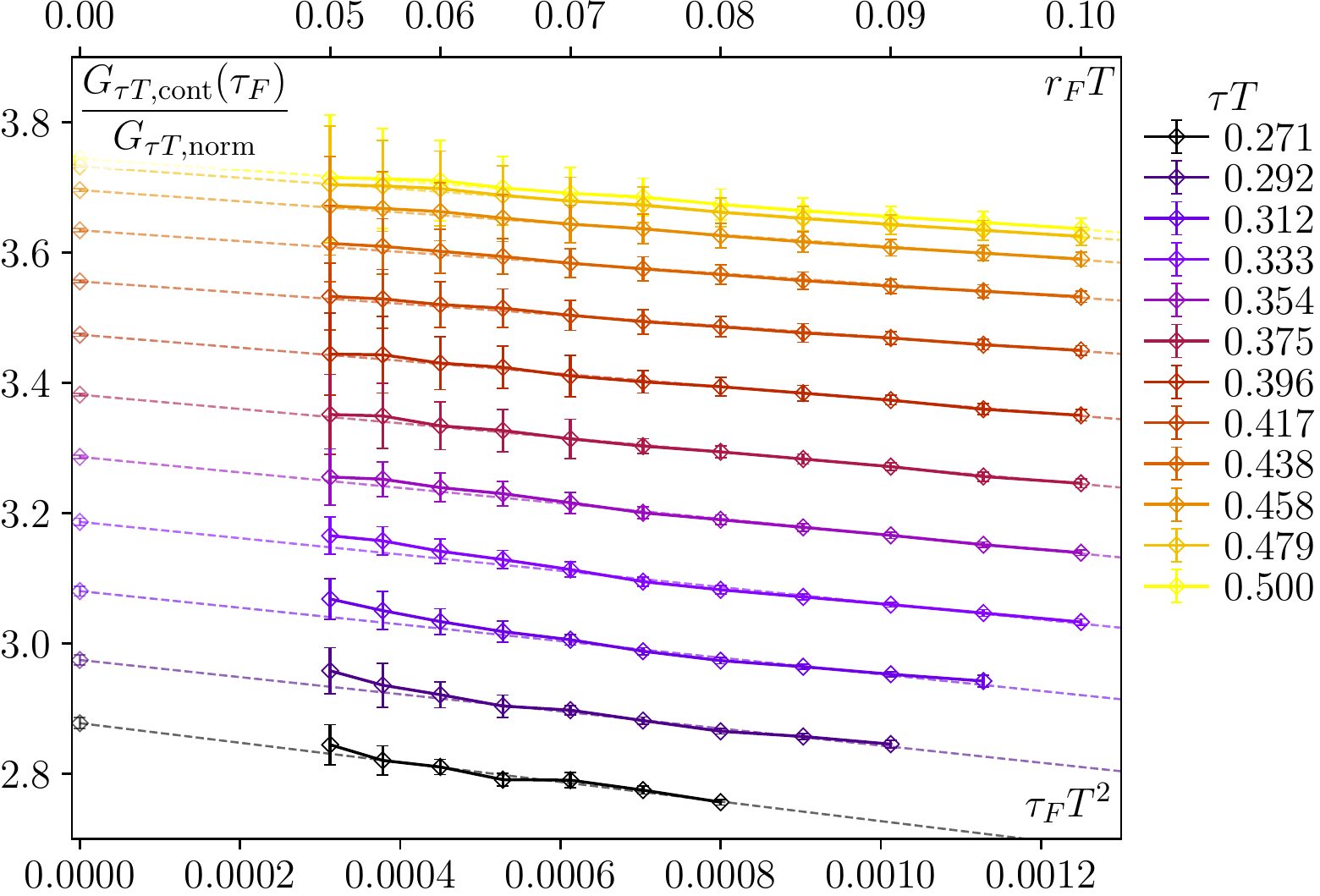}
}
\caption{Extrapolate the continuum correlators to zero flow time using linear ansatz at different $\tau T$.}
\label{tto0}
\end{figure}

\begin{figure}[h]
\centerline{
\includegraphics[width=0.5\textwidth]{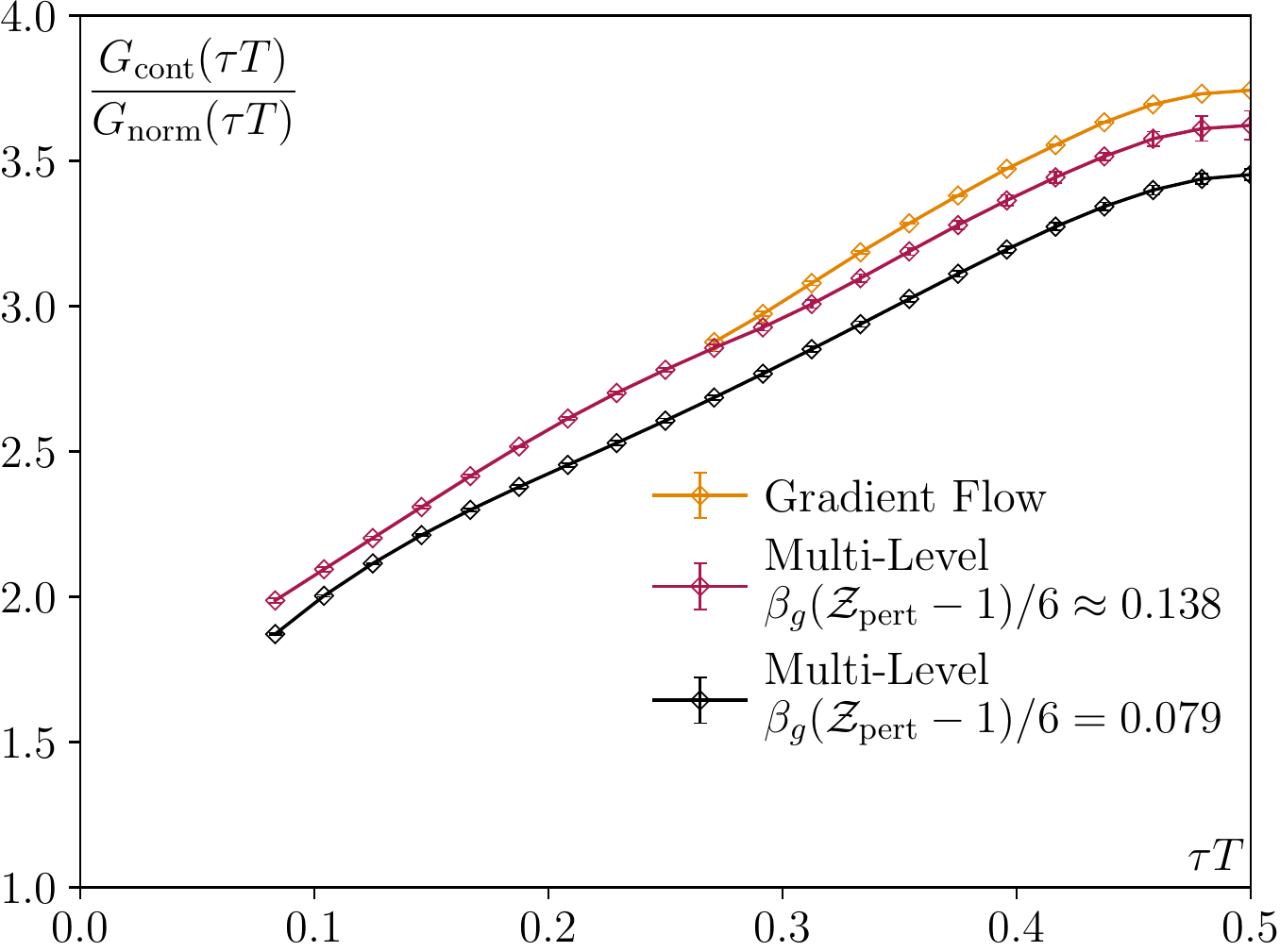}
}
\caption{A comparison of the color-electric correlator obtained from gradient flow (\textit{orange}) and the correlator from the multi-level algorithm \cite{Francis:2015daa} (\textit{black}) with updated renormalization factor~\cite{Christensen:2016wdo} (\textit{red}). The flowed correlator has been double extrapolated($a\rightarrow 0, \tau_F\rightarrow 0$) and the correlator from the multi-level method has been taken to the continuum limit.}
\label{comp}
\end{figure}

In this section we show how we extrapolate the continuum correlators obtained above to zero flow time. 
The continuum extrapolated correlators seem to depend linearly on flow time, which we show for some selected $\tau T$ in figure \ref{tto0}. For the extrapolation we use the Ansatz 
\begin{equation}
G(\tau T, \tau_F)=c(\tau T)\cdot \tau_F+b(\tau T, \tau_F=0),
\end{equation} 
which is customary for actions with $\mathcal{O}(a^2)$ cutoff-effects. Here $c$ and $b$ are fit parameters. We finally take $\tau_F\rightarrow 0$ and obtain the correlator at both the continuum and zero flow time limit, which we show in figure~\ref{comp}.

 
In figure.\ref{comp} we compare the correlator obtained by gradient flow after takinig the $a\rightarrow 0$ and $\tau_F\rightarrow 0$ limit with one from \cite{Francis:2015daa}, which was obtained by the multi-level algorithm. We performed a new continuum extrapolation of the original multi-level correlator where we use the corrected renormalization factor~\cite{Christensen:2016wdo} (the one used in \cite{Francis:2015daa} had an algebraic error).
In the previous multilevel study we used lattices up to $192^3\times 48$ and a reliable continuum extrapolation was possible down to $\tau T\sim 0.1$, while in the gradient flow study so far we have used up to $120^3\times 30$ lattice allowing for an extrapolation only for $\tau T\sim 0.25$. However, the behavior of the correlators is very similar. The difference between the multilevel and flow results can be attributed to the uncertainty in the renormalization of the multilevel results while the flow results provide a non-perturbatively renormalized correlator in the continuum.

\section{Summary}

We have calculated a color-electric correlator on four different isotropic lattices at $T\approx 1.5T_c$ in the quenched approximation under gradient flow. We found that the gradient flow technique significantly improves the signal of the correlator on the lattice. With our data sets we are able to perform both a continuum and a zero flow time extrapolation. We found that an Ansatz linear in $1/N_{\tau}^2$ and in $\tau_F$ respectively is suitable for the extrapolations. The overall shape of the flowed correlator in the $a\rightarrow 0$ and $\tau_F\rightarrow 0$ limit agrees with its multi-level counterpart in the $a\rightarrow 0$ limit. This work affirms the validity of the gradient flow technique when performing the continuum and flow time limit and provide a method for extending these studies to full QCD. 

\section{Acknowledgement}
This work is supported by the Deutsche Forschungsgemeinschaft (DFG, German Research Foundation)-Project number 315477589-TRR 211. The computations in this work were performed on the GPU cluster at Bielefeld University.

\end{document}